\documentstyle[11pt,epsfig]{article}

\newcounter{savefig}
\newcommand{\alphfig}{\setcounter{savefig}{\value{figure}}%
\setcounter{figure}{0}%
\renewcommand{\thefigure}{\mbox{\arabic{savefig}\alph{figure}}}}

\newcommand{\be}{\begin{equation}}
\newcommand{\ee}{\end{equation}}
\newcommand{\beqn}{\begin{eqnarray}}
\newcommand{\eeqn}{\end{eqnarray}}

\newcommand{\Pma}{I\!\!P}

\begin{document}
\begin{center}
\begin{Large}
\boldmath
{\bf A Numerical Estimate of the Small-$k_T$ Region in the BFKL Pomeron }\\
\end{Large}
\unboldmath
\vspace{0.5cm}
J.\,Bartels$^1$, H.\,Lotter$^1$, and M.\,Vogt$^2$\\
{\it $^1$ II. Institut f\" ur Theoretische Physik, Universit\" at Hamburg
\footnote
        {Supported by Bundesministerium f\"ur Forschung und
        Technologie, Bonn, Germany under Contract 05\,6HH93P(5) and
        EEC Program "Human Capital and Mobility" through Network
        "Physics at High Energy Colliders" under Contract
        CHRX-CT93-0357 (DG12 COMA)} 
\\ 
$^2$ Deutsches Elektronen Synchrotron DESY, Hamburg} \\ 

hep-ph/9511399
\end{center}
\vspace{2.0cm}
\begin{abstract}
\noindent
A computer study is performed to estimate the influence
of the small-$k_T$ region in the BFKL evolution equation.
We consider the small-x region of the deep inelastic structure
function $F_2$ and show that the magnitude of the small-$k_T$ region
depends on $Q^2$ and $x_B$. We suggest that the width of the
$\log k_T^2$-distribution in the final state may 
serve as an additional footprint
of BFKL--dynamics. For diffractive dissociation it is shown that the
contribution of the infrared region is large - even for large $Q^2$.
This contribution becomes smaller only if restrictions on the final
state are imposed. 
\end{abstract}
\vspace{1cm}
\noindent
{\bf 1.)} Deep inelastic electron proton scattering at HERA has put strong
emphasis on small-x physics. Both the observed rise of $F_2$ at small x  
and the discovery of the large rapidity gap events have
raised the question of how to connect these observations with the
perturbative Balitskii, Fadin, Kuraev, Lipatov (BFKL) 
Pomeron \cite{BFKL} and the (nonperturbative) soft Pomeron \cite{DL} seen in 
hadron hadron scattering at high energies.\\ \\
In the context of the deep inelastic structure functions the
BFKL Pomeron suggests the possibility of calculating, for a not too small
momentum scale $Q_0^2$, the x-dependence of the input gluon
distribution which then enters the DGLAP \cite{DGLAP} evolution
equations. The main result is the power-like increase
$xg(x,Q^2) \sim (1/x)^{\omega_{{\rm BFKL}} }$ with $\omega_{{\rm BFKL}} =
\frac{N_c \alpha_s}{\pi} \, 4 \log 2 \approx 0.5$. Such an increase
is consistent with the data, but the preferred power is slightly
smaller:  for the gluon distribution
$xg(x,Q^2)$ the observed power lies in the
interval $\omega_{{\rm exp}} = 0.3 ...0.4$ \cite{H1}, and for $F_2$ one finds
a power in the range 0.2...0.4. A recent analysis \cite{H1} leads to the
conclusion that the experimental results are compatible with the BFKL 
interpretation, but the evidence is not yet compelling. \\ \\
Unfortunately, numerical predictions based 
upon the BFKL approximation have several principal uncertainties.
Most important is the observation that, as a result of the diffusion of 
$\log k_T^2$, the internal transverse momenta of the BFKL ladders may become 
arbitrarily small, and the contribution from this region of phase space 
cannot be extracted reliably within the leading logarithmic approximation. 
One expects that in this region higher order corrections will be very 
important, and eventually nonperturbative effects will come into play.
A convenient way of taking into account higher order perturbative QCD effects 
is the replacement of the fixed coupling constant by the running $\alpha_s$. 
This {\it ad hoc} procedure, however, is somewhat arbitrary: 
the BFKL ladders represent the leading logarithmic approximation, and the
QCD coupling constant is frozen at the scale  set by the external particles
(e.g. the virtuality of the scattering gluons). The introduction of the
running coupling is equivalent to taking into account a certain 
subset of nonleading corrections, but is clearly incomplete and misses 
out other important corrections.
 Moreover, the presence of the Landau pole in $\alpha_s(k_T^2)$ 
requires an infrared cutoff which again represents a model dependent choice.
These uncertainties, clearly, will be most severe if, in a given application,
the BFKL Pomeron receives a large contribution from the infrared region.
Therefore, a first estimate of the reliability of a BFKL prediction
is a numerical study of the magnitude of the infrared region in the BFKL 
evolution (with fixed coupling $\alpha_s$). If it turns out that the 
contribution of the infrared region is small, there seems to be no strong
need of introducing the running coupling, and the BFKL calculation can be
expected to be rather safe.\\ \\
From this point of view the safest place to test the BFKL prediction
at HERA is the inclusive measurement of forward jets \cite{Hot,BDDGW}:
here the BFKL Pomeron has large momentum scales at both of its
 ends (the large
photon mass and the transverse momentum of the observed jet). Consequently,
the diffusion into the infrared region is expected to be small. This 
has been confirmed in a numerical analysis of the BFKL evolution
\cite{BaLo}. The deep inelastic structure function $F_2$, on the other hand,
is not a particularly suitable place to test the BFKL Pomeron. The large 
momentum scale of the photon mass provides a justification for applying 
perturbation theory, but the proton at the other end of the BFKL ladder 
pulls the momentum scale down to small momentum, and during the BFKL 
evolution the infrared region may contribute a quite significant amount.
In order to obtain a more precise estimate of the magnitude of this
``dangerous`` region again a computer analysis is needed, quite analogous 
to \cite{BaLo}. Another place at HERA where the BFKL Pomeron might
play an important role is the diffractive dissociation of the photon.
Apart from analytic considerations \cite{BW,BLW} so far no estimate 
of the contribution of small $k_T^2$ to the BFKL evolution has been 
performed.  
 \\ \\
{\bf 2.)} In this letter we present the results of a numerical calculation of 
the BFKL Pomeron for $F_2$, following the strategy of \cite{BaLo}: 
writing the BFKL equation (with fixed $\alpha_s$) as an evolution equation 
in $y=\log 1/x$, we start
the evolution at some initial value $x_0$, evolve in $y$ and compute, as
a function of $y$, the distribution in the logarithm of the transverse 
momentum, $\log k_T^2$. In our choice of the 
physical parameters we  follow \cite{AKMS}: we take $x_0=10^{-2}$ and 
evolve down to the HERA value of $x=10^{-4}$.
The initial $k_T$-distribution $f_0$ is calculated from a  
recent GRV-parameterization \cite{GRV} of the gluon
density function of the proton.
We model the nonperturbative region
below $Q_0^2=1.5 \, \mbox{GeV}^2$ by using 
the ansatz \cite{AKMS}
\beqn
f_0(x,k_T^2)=\mbox{C} \frac{k_T^2}{k_T^2+Q_0^2}.
\label{formf}
\eeqn
The constant C has been chosen in such a way that
continuity of $f_0$ at $Q_0^2$ is guaranteed.
At the end of the evolution we convolute the gluon distribution with the
quark loop \cite{loop}, where the quarks are taken to be massless.
A first analysis in this direction 
but with the ansatz (\ref{formf}) taken for the whole
$k_T^2$-range was presented in \cite{fhs}.
\\ \\
Results are shown in Fig.1. The straight line in the center denotes,
as a function of $y$, the mean value, whereas the curved lines to the left
and the right mark the mean square deviation from the mean value.
We have considered two different
values for $Q^2$: $10 \, \mbox{GeV}^2$ (Fig. \ref{f1a})
and $100 \, \mbox{GeV}^2$ (Fig. \ref{f1b}). 
One observes that a 
large value of $Q^2$ ``drags``
the distribution towards larger values, i.e. away from the dangerous infrared
region. Nevertheless, at the upper end the mean value stays somewhat below
the photon mass $Q^2$. Increasing, on the other hand, the available 
rapidity interval,
one 
makes the integration region broader and, hence, increases the influence
of the infrared domain. In summary, one feels that the application of the
BFKL Pomeron for $F_2$, in particular for the lower $Q^2$ value, is somewhat
uncertain. Most likely, nonleading corrections to the BFKL Pomeron cannot
be neglected in this region; the calculation of these corrections therefore
appears to be quite urgent.
\\ \\
In order to provide further evidence for BFKL dynamics and, in particular, to
discriminate between BFKL and DGLAP dynamics, it is useful to look into
the distribution of transverse energy in the final state. Some time ago
it has been argued \cite{Durh1} that a measurement of the 
mean transverse energy $<E_T>$ of produced jets could be used to discriminate 
between a ``BFKL`` and a ``DGLAP`` final state. Namely, 
because of the absence of the strong $k_T$-ordering in the BFKL case, one
expects a larger mean value $<E_T>$ than in the DGLAP-case. Moreover,
the dependence upon $x$ is quite different in both cases. For the
BFKL case the general trend can be read off from the distributions shown in 
Fig.\ \ref{f1a} - \ref{f1b}.
For simplicity we study the $< E_T >$ of the t-channel gluons
which is not exactly the same as that of the produced jet. One 
should however expect that, qualitatively, both variables show
the same behavior. 
%
We focus on the center in pseudorapidity in the hadronic center
of mass system and study
the mean value in $\log k_T^2$ and the mean spread to the right and the left.
By lowering $x_B$ (i.e. increasing the length in rapidity) one observes 
an increase of the width; for the  mean value there is no visible motion.
In order to study these qualitative expectations in more detail, we
show, in Fig.\ \ref{f2a}, the distribution in $\log k_T^2$
 in the center of the rapidity
interval between the photon and the proton, varying the overall length in
rapidity. One sees that the maxima do not move when rapidity 
increases whereas the width clearly increases with growing
rapidity. Simple analytic considerations in fact show that the 
maximum does not move. For the width the random walk mechanism in
 the BFKL Pomeron leads to the prediction: 
\beqn
\label{width}
\Delta( \log k_T^2 ) &=&
\sqrt{ \frac{N_c \alpha_s}{\pi} 28 \zeta(3) \, y} \; ,\\
y&=&\frac{(\alpha_1+\frac{1}{2} \log \frac{1}{x_B} \frac{k_{J,T}^2}{Q^2})
        (\alpha_2-\frac{1}{2} \log \frac{x_B}{x_0^2} \frac{k_{J,T}^2}{Q^2})}
       {\alpha_1+\alpha_2+\log \frac{x_0}{x_B}}  . \nonumber
\eeqn
Here $\alpha_1$ and $\alpha_2$ are determined by the width of the boundary 
distributions at $x_B$ and $x_0$. For $k_{J,T}^2$ we choose a value 
of $2 \mbox{GeV}^2$ which is an estimate based on the mean $k_T^2$
displayed in Figs.\ \ref{f1a} - \ref{f1b}.
Fig.\ \ref{f2b} shows the growth of the width with 
increasing rapidity. For comparison we display the analytic 
prediction (\ref{width}) for $Q^2 = 100 \mbox{GeV}^2$. 
For small $x_B$ the 
analytic curve approaches the numerical one.  
We therefore 
conclude that the behavior of the width (in $\log k_T^2$ )
may be a good signal for
BFKL dynamics in the final state.
In \cite{AKMS,Exp4} the variable  
$E_T$ has been used to study ``footprints`` of the BFKL Pomeron.
From the $\log k_T^2$ - distributions presented in Fig.\ \ref{f2a}
we have calculated the distributions in
$E_T = k_T = \exp \,\frac{1}{2}\log k_T^2$.
The growth of 
$<E_T>$ shown in Fig.\ \ref{f3b}, indeed, presents a
characteristic BFKL signal.  The analytic prediction
for this observable is
\beqn
< E_T > &=& \exp \left[ \frac{N_c \alpha_s}{\pi} \frac{7}{2} \zeta(3) y
+ \frac{1}{2} \xi \right] ,\\
\xi &=& \frac{<\log k_{1,T}^2> (\alpha_2
             -\frac{1}{2} \log \frac{x_B}{x_0^2} \frac{k_{J,T}^2}{Q^2})
            +<\log k_{2,T}^2> (\alpha_1                             
               +\frac{1}{2} \log \frac{1}{x_B} \frac{k_{J,T}^2}{Q^2})}
           {\alpha_1+\alpha_2+\log \frac{x_0}{x_B}}.   \nonumber
\eeqn
It depends on the mean $\log k_{i,T}^2$ 
($i=1,2$) of the two boundary contributions.
The difference seen in the $x_B$-dependence of $<\log k_T^2 >$ 
(Fig.\ \ref{f2a}) and $<E_T>$
(Fig.\ \ref{f3b}) indicates that the
latter quantity already feels 
the increase of the width. From the
point of view of BFKL predictions, however,
the variable $\log k_T^2$ and its width seem to be more natural. \\ \\
As far as comparison with experiment is concerned,  
for the mean value $<E_T>$ of produced jets, there is
evidence \cite{Exp4} that the data are, in fact, closer to the BFKL-type
distribution than to DGLAP; however, in the comparison with the data,
the ``theoretical predictions`` are presented by two different Monte Carlos
which are neither pure BFKL nor DGLAP. It is, on the other hand, quite
clear that the hadronization effects will lead to changes of the analytical
predictions. For the width an analysis along this line has not yet been done.
\\ \\
{\bf 3.)} Another place where the BFKL Pomeron
 may potentially become important
is the measurement of the diffractive dissociation of the photon.
${\it A\,\,priori}$, it is not clear whether the Pomeron which is responsible 
for the rapidity gap is the same as the one seen in hadron hadron 
scattering or whether it is more close to the hard Pomeron seen in
$F_2$. Following a straightforward partonic picture of this process,
one is lead to the conclusion that the integrated cross section (at fixed
mass $M$ of the diffractive system) should be a superposition of both types 
of the Pomeron, the hard and the soft one. The process is most
conveniently discussed in the proton rest frame: at small x, the 
photon splits 
into a $q\bar{q}$ pair, and the lifetime of this fluctuation is of the order
$1/2M_px$. In the simplest case the final state of the photon consists just 
of this quark-antiquark pair which scatters elastically off the proton.
The dynamics of the elastic scattering now depends crucially upon the
kinematic configuration of the $q\bar{q}$-pair. If the transverse momenta
are small and the mass $M$ of the system is of the order of $\sqrt{Q^2}$, the
longitudinal momenta should be rather asymmetric , i.e. one of the quarks 
carries almost all the longitudinal momentum of the photon, whereas the
momentum fraction of the other one is small.
 In the $\gamma^*-\Pma$ rest system
the produced quark and antiquark are almost collinear with the proton
direction (``aligned jet model`` \cite{AJM}). Since the elastic 
scattering of a
quark with small virtuality is expected to proceed via the soft Pomeron, this
final state configuration will be dominated by the soft Pomeron and its 
characteristic energy dependence. In addition to this configuration, 
however,  there is also the final state with larger transverse momenta. In 
this case one expects the elastic scattering of the $q\bar{q}$-system to 
proceed via the exchange of the hard Pomeron, e.g. a perturbative BFKL 
ladder. A signal of this would be a steeper energy dependence of the cross 
section. Combining these two possibilities, one expects to see both the
hard and the soft Pomeron, each of them being associated with specific
final state configurations. As to the question which of them will be 
dominant, there seems to be, at least within this picture, a slight 
preference for the soft configuration: a quark with a large transverse 
size has a larger cross section than a small-size object. 
On the other hand, the appearance of the hard Pomeron which at large energies
grows stronger than the soft one enhances the cross section of the hard
diffractive final state. Clearly a careful study of QCD models based upon
the BFKL Pomeron will be a valuable way of studying this competition 
between the hard and the soft component. \\ \\
In order to discuss the experimental situation we need a few formulae.
Arguments based upon triple Regge analysis \cite{BCSS,IP,Exp1} suggest to 
write the cross section in the following form:
\beqn
\frac{d^4\sigma}{dxdQ^2dx_{\Pma} dt}=
\frac{4\pi \alpha^2}{xQ^2}(1-y+\frac{y^2}{2})
F_2^{\Pma} (\beta,Q^2,t) (\frac{1}{x_{\Pma}})^{2 \alpha(t)-1} 
\frac{\beta_p^2(t)}{16 \pi},
\label{f2d}
\eeqn
where $x_B=\frac{Q^2+M^2-t}{Q^2+W^2}$, $\beta=x/x_{\Pma} =
 \frac{Q^2}{Q^2+M^2-t}$,
and we have neglected the longitudinal part of the Pomeron 
structure function.
The soft Pomeron \cite{DL} requires the exponent of $1/x_{\Pma}$ to have the 
value $2 \alpha(0) -1 =1.17$, whereas the hard Pomeron belongs to 
$2\alpha(0)-1 >1.4$ (if one uses the power 0.2 for the hard Pomeron at 
$20 \, \mbox{GeV}^2$), or even $2\alpha(0)-1 \approx 2$ for the BFKL Pomeron.
Experimental values for the exponent of
$\frac{1}{x_{\Pma}}$  are \cite{Exp1,Exp2,Exp3}:
\begin{center}
\begin{tabular}{|c|c|c|c|}   \hline
 H1 & 1.19        &  $\pm 0.06$       &  $\pm 0.07$        \\  
 \hline
 ZEUS & 1.30        &  $\pm 0.08$        
 & $\stackrel{\scriptstyle+0.08}{\scriptstyle -0.14}$     \\ 
  \hline
 ZEUS & 1.47        &  $\pm 0.03$       & 
 $\stackrel{\scriptstyle+0.14}{\scriptstyle -0.10}$      \\ 
  \hline
\end{tabular}
\end{center}
The third value is based upon a new method of defining the nondiffractive
background. Obviously, the first value is close to the soft Pomeron, i.e.
in most of the events the final state is such that the soft Pomeron 
dominates.
The third value, on the other hand, seems to favor the hard Pomeron. 
\cite{Exp1,Exp2} also present evidence for the factorization in (\ref{f2d}):
 within the errors, the power of $1/x_{\Pma}$ does
 not vary with $Q^2$ or $\beta$. \\ \\
Perturbative QCD calculations of diffractive dissociation are contained in 
\cite{Ry,NN,WL}; a consistent BFKL-type calculation has been presented
in \cite{BW}. The latter one, being based upon the BFKL approximation, also
includes small transverse momenta for the produced diffractive system
consisting of $q\bar{q}+gluons$. Therefore
it allows, in particular, to answer the question which transverse momenta
the BFKL Pomeron predicts for the final state. Analytic estimates based upon
the results of \cite{BW} show that the preferred $k_T$ value at the upper end
of the BFKL Pomeron in diffractive DIS dissociation is rather low. In other
words, in this process the BFKL Pomeron seems to get its main contribution 
from the region of small $k_T$ where higher order corrections are expected to
be essential. Their main effect will be the lowering of the exponent
$\omega_{{\rm BFKL}}$. Since we do not know yet the corrections to BFKL,
we are presently unable to make a quantitative prediction.\\ \\
{\bf 4.)} Starting from the cross section formula
 for the diffractive production
of $q\bar{q}$ pairs (at $t=0$) we have
 performed a numerical analysis of the
BFKL Pomeron, very much in the same spirit as we did for $F_2$.
 The coupling 
to the proton is done in the same way as in $F_2$ (see above).
In Figs.\ \ref{f4a} -- \ref{f4d}
we show, as a function of rapidity, the distribution of the transverse
momentum inside the BFKL ladder. We begin with the
 integrated (over invariant
mass $M$ and transverse momenta of the outgoing quarks) cross
 section in
Figs.\ \ref{f4a} and \ref{f4b}. Most striking is the trend of the mean value
to be small at the upper end (i.e. the coupling to
 the $q\bar{q}$): increasing
the photon mass $Q^2$ seems to pull the upper value
 to the right, but it does
not exceed the hadronic scale at the lower
 end (i.e. the coupling to the
proton). This fully confirms the findings of \cite{BLW},
 and we are lead 
to the conclusion that the BFKL Pomeron in diffractive dissociation 
gets its main contributions from a region where the corrections
 should be 
large.
Based upon the observed behavior of $F_2$ we expect
 that these corrections
will mainly lead to a lowering of the increase in $1/x_{\Pma}$, i.e. to the
change from the hard Pomeron to the soft one. Next we undo the 
integration over the final state: at fixed values for
$Q^2$ ($20 \, \mbox{GeV}^2$) we plot the cross section
 (integrated over the invariant
mass $M$) for different 
values of the transverse momenta of the outgoing quarks:
 Figs.\ \ref{f4c} and \ref{f4d} ($p_T^2= 1 \, \mbox{GeV}^2$ and
$5 \, \mbox{GeV}^2$ resp.) show the clear
 trend that a large $p_T$ pulls the scale
at the upper end of the Pomeron to the right and thus makes the Pomeron
harder.
This is consistent with the calculations \cite{Ry2,BFGMS} of the 
cross section for the production of heavy vector mesons from 
longitudinal photons: in the coupling of the pomeron to the 
$q \bar{q}$ - state large $p_T^2$ dominate and one expects the hard 
pomeron to become important. Recent HERA data \cite{HVM} seem to
support this prediction. 
\\ \\
This result has implications for the applicability of
 perturbative models
to the process of diffractive dissociation. In earlier
 models \cite{Ry,NN,WL}
it was assumed that the final states had to be hard. The model in 
\cite{BW} which has been used in this letter, has the advantage that
it is well-defined also in the small-$k_T$ region.
 The analysis, however,
has lead to the conclusion that the hard final states are
 only a minor part
of the cross section, and for a reliable calculation higher order
corrections are unavoidable. A first attempt to model higher order
effects has been made recently \cite{Wue}: by replacing the fixed power
of the BFKL Pomeron by a variable one (being a function of the
scale of the $q\bar{q}$ pair), an effective transition from the 
hard to the soft Pomeron has been included. Further work in this direction
appears to be very promising.
\\
As an experimantal consequence of our study we expect that events with
hard $q\bar{q}$ - final states will get a larger contribution
from the hard pomeron. As a first guideline one may trigger 
on events with large $p_T^2$ (in the $\gamma^{\ast} - p $ center of 
mass system).
A more accurate estimate of the hardness of the pomeron 
may be $p_T^2(Q^2+M^2)/M^2$ which, in the aligned jet model,
would correspond to the virtuality of one of the two quarks
in which the photon dissociates.
\\ \\
{\bf 5.)} More generally, the behavior of the BFKL Pomeron in diffractive
dissociation may shed some light on the way in which the unitarization of the
BFKL Pomeron works. It is generally believed that among the next-to-leading
order corrections to the BFKL Pomeron those with a larger number of t-channel
gluon lines ($n>2$) will play the most important role in unitarizing the
BFKL approximation. The cross section for diffractive dissociation
(as discussed in, e.g., \cite{BW}) constitutes a measurable part of these
corrections. Consequently, a measurement of the cross section and the
comparison with the calculation can be used to estimate the 
magnitude of unitarity corrections, and a detailed study of models
can teach us how unitarization will work. As to the discussion
presented in this letter, the main conclusion in this direction seems to be
that unitarization of the BFKL Pomeron begins mainly in the small-$k_T$
region.\\ \\ 
\newpage
\newpage
\section*{Figure captions}
\begin{description}
\item
Fig.\ \ref{f1a} : Mean value and root mean square deviation of the
distribution of $\log k_T^2$ inside the BFKL--evolution for
$F_2$. The values
of $x_B$ and $Q^2$ are $10^{-4}$ and $10 \,\mbox{GeV}^2$. 
The variable $x$ gives the position on the BFKL ladder at which these
quantities are measured.
\\
\item
Fig.\ \ref{f1b} : The same as in Fig.\ \ref{f1a} but with 
$x_B = 10^{-4}$ and $Q^2 = 100 \,\mbox{GeV}^2$.  
\\
\item
Fig.\ \ref{f2a} : Normalized distribution of $\log k_T^2$
 inside the BFKL--evolution for
$F_2$ with $Q^2=100 \,\mbox{GeV}^2$ and three different values of $x_B$,
 measured at a rapidity value of $\frac{1}{2} \log  \frac{10^{-2}}{x_B}$.
Values of $x_B$ are $5 \cdot 10^{-3}$ (the curve with the highest peak),
$10^{-3}$ and $5 \cdot 10^{-4}$ (the curve with the lowest peak).
\\
\item
Fig.\ \ref{f2b} : Dependence of the the root mean square deviation
of the distributions of Fig.\ \ref{f2a} on $x_B$ for  
$Q^2 = 100$,  30 and 15 
$\mbox{GeV}^2$ (from top to bottom).
The dotted line shows the analytic prediction
eq. (\ref{width}) for $Q^2 = 100$ $\mbox{GeV}^2$.
\\
\item
Fig.\ \ref{f3b} : Dependence of the mean transverse energy
$E_T=\exp \frac{1}{2} \log k_T^2$
of the 
distributions of Fig.\ \ref{f2a} on $x_B$ shown for values of
$Q^2 = 100$,  30 and 15 $\mbox{GeV}^2$ (from top to bottom).
\\
\item
Fig.\ \ref{f4a} : Mean value and root mean square deviation of the 
$ \log k_T^2 $ distribution inside the BFKL--evolution in 
totally inclusive
diffractive dissociation. Values of $x_B$ and $Q^2$ are $10^{-4}$ and 
$10 \, \mbox{GeV}^2$.  
\\
\item
Fig.\ \ref{f4b} : The same as in Fig.\ \ref{f4a} but with 
$x_B = 10^{-4}$ and $Q^2 = 100 \, \mbox{GeV}^2$.  
\\
\item
Fig.\ \ref{f4c} : Mean value and root mean square deviation of the  
$ \log k_T^2 $ distribution inside the BFKL--evolution in 
the M -- integrated 
diffractive dissociation with high $p_T$-quarks in the 
final state. We have taken $x_B = 10^{-4}$, $Q^2 = 20 \,\mbox{GeV}^2$ 
and $p_T^2 = 1 \,\mbox{GeV}^2$.  
\\
\item
Fig.\ \ref{f4d} : The same as Fig.\ \ref{f4c} with
 $x_B = 10^{-4}$, $Q^2 = 20 \,\mbox{GeV}^2$ 
and $p_T^2 = 5 \,\mbox{GeV}^2$. 
\end{description}
\clearpage
\newpage
\setcounter{figure}{1}
\alphfig
\setlength{\unitlength}{1cm}
\begin{figure}[t] 
\begin{center}
\begin{picture}(8,8)(0,0)
\put(-6.55,-18.2){\epsfig{file=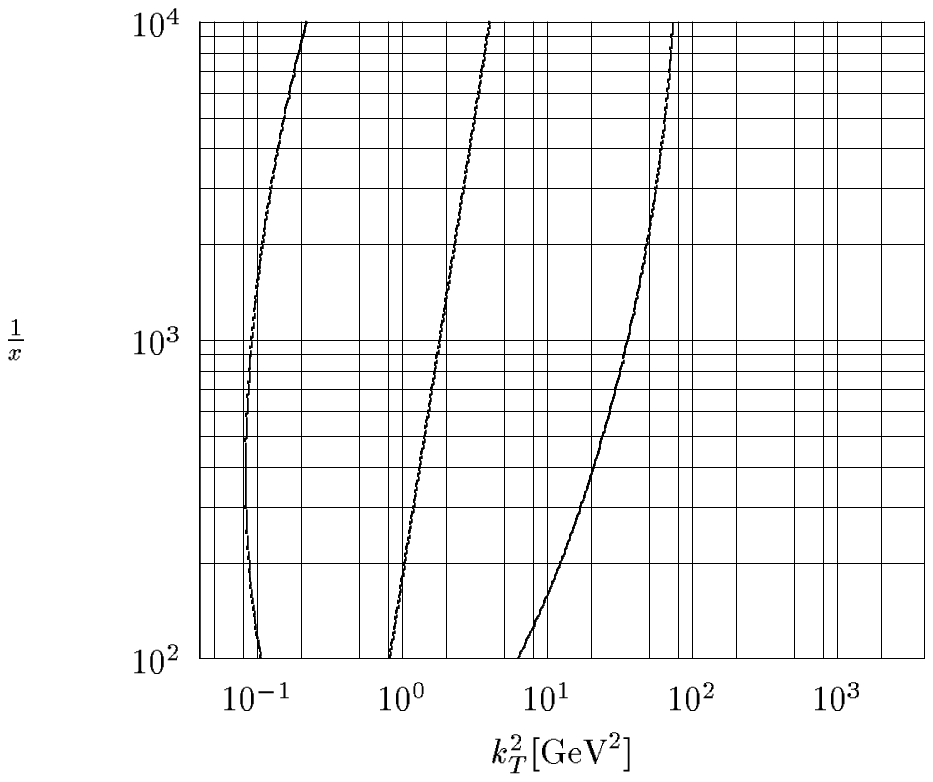,height=31.5cm,width=20.6cm}}
\end{picture}
\end{center}
\caption{}
\label{f1a}
\end{figure}
\begin{figure}[b] 
\begin{center}
\begin{picture}(8,8)(0,0)
\put(-6.55,-18.2){\epsfig{file=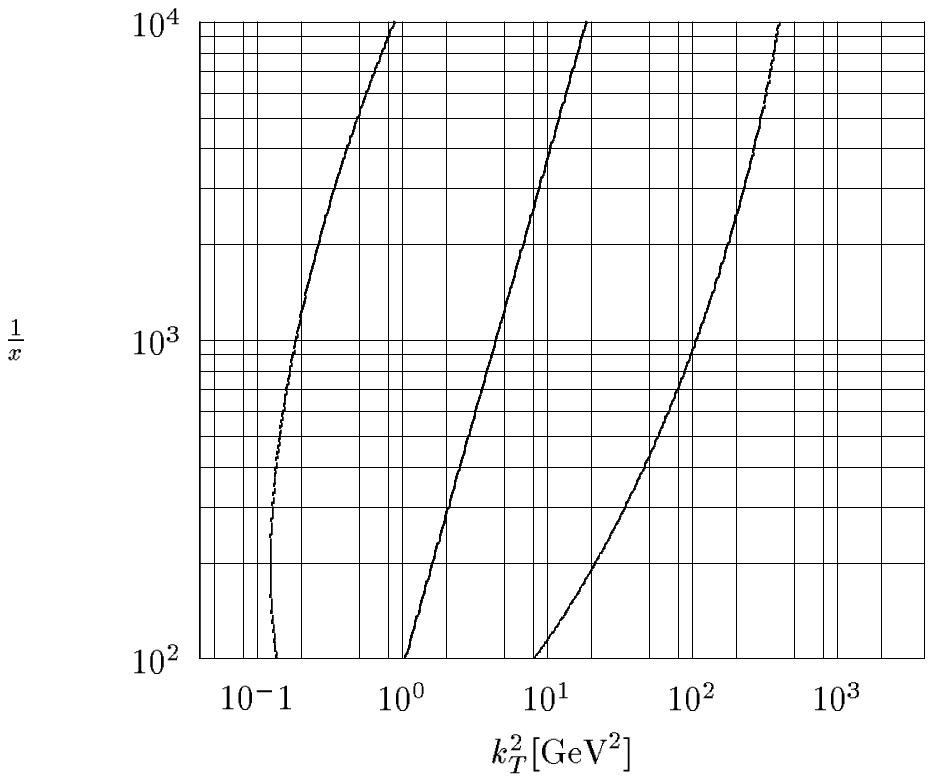,height=31.5cm,width=20.6cm}}
\end{picture}
\end{center}
\caption{}
\label{f1b}
\end{figure}
\clearpage
\newpage
\setcounter{figure}{2}
\alphfig 
\begin{figure}[t] 
\begin{center}
\begin{picture}(8,8)(0,0)
\put(-6.55,-18.2){\epsfig{file=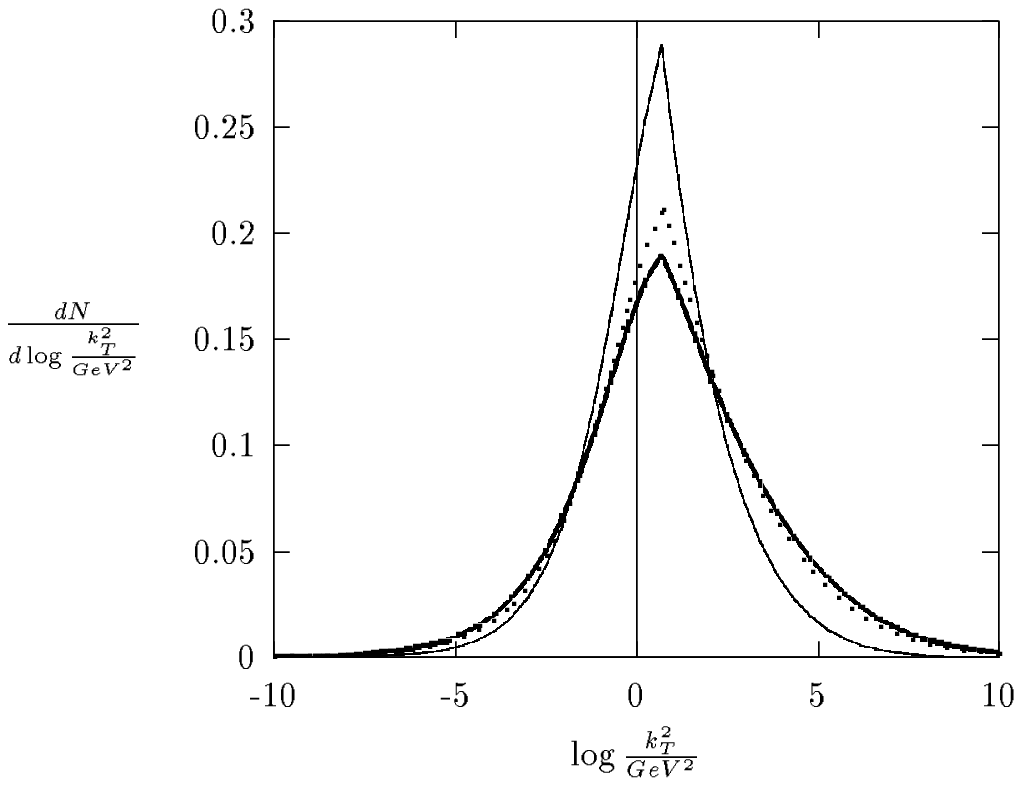,height=31.5cm,width=20.6cm}}
\end{picture}
\end{center}
\caption{}
\label{f2a}
\end{figure}
\begin{figure}[b] 
\begin{center}
\begin{picture}(8,8)(0,0)
\put(-6.55,-18.2){\epsfig{file=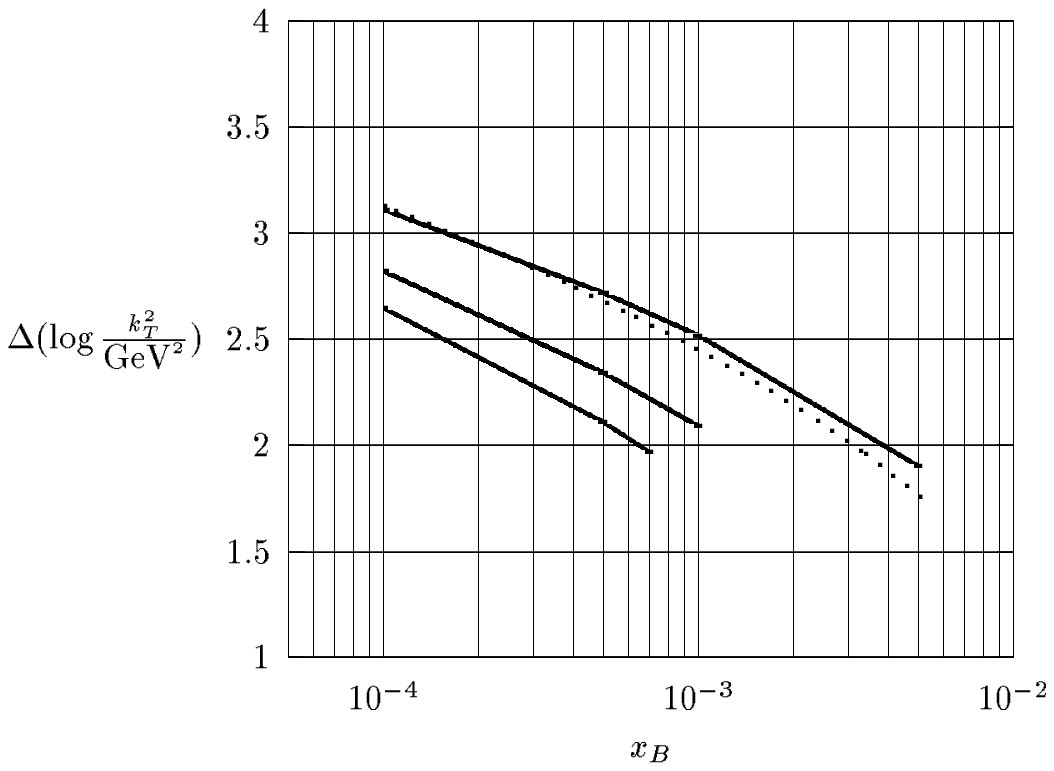,height=31.5cm,width=20.6cm}}
\end{picture}
\end{center}
\caption{}
\label{f2b}
\end{figure}
\clearpage
\newpage
\begin{figure}[h] 
\begin{center}
\begin{picture}(8,8)(0,0)
\put(-6.55,-18.2){\epsfig{file=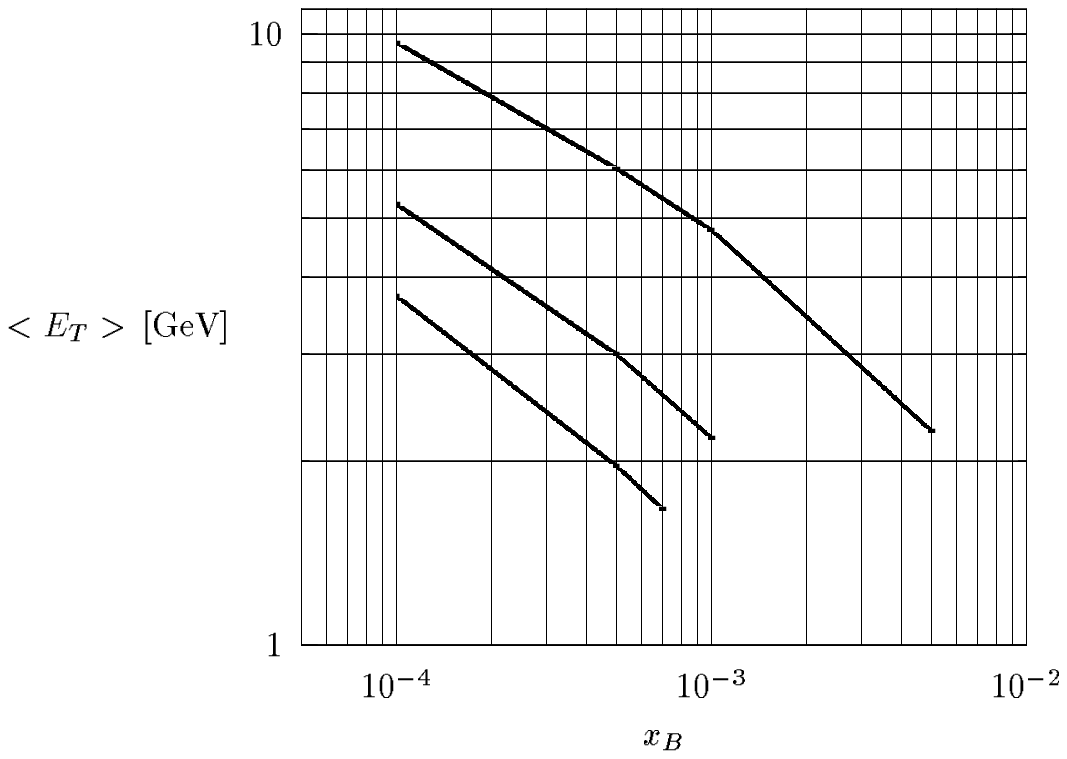,height=31.5cm,width=20.6cm}}
\end{picture}
\end{center}
\caption{}
\label{f3b}
\end{figure}
\clearpage
\newpage
\setcounter{figure}{3}
\alphfig
\begin{figure}[t]
\begin{center}
\begin{picture}(8,8)(0,0)
\put(-6.55,-18.2){\epsfig{file=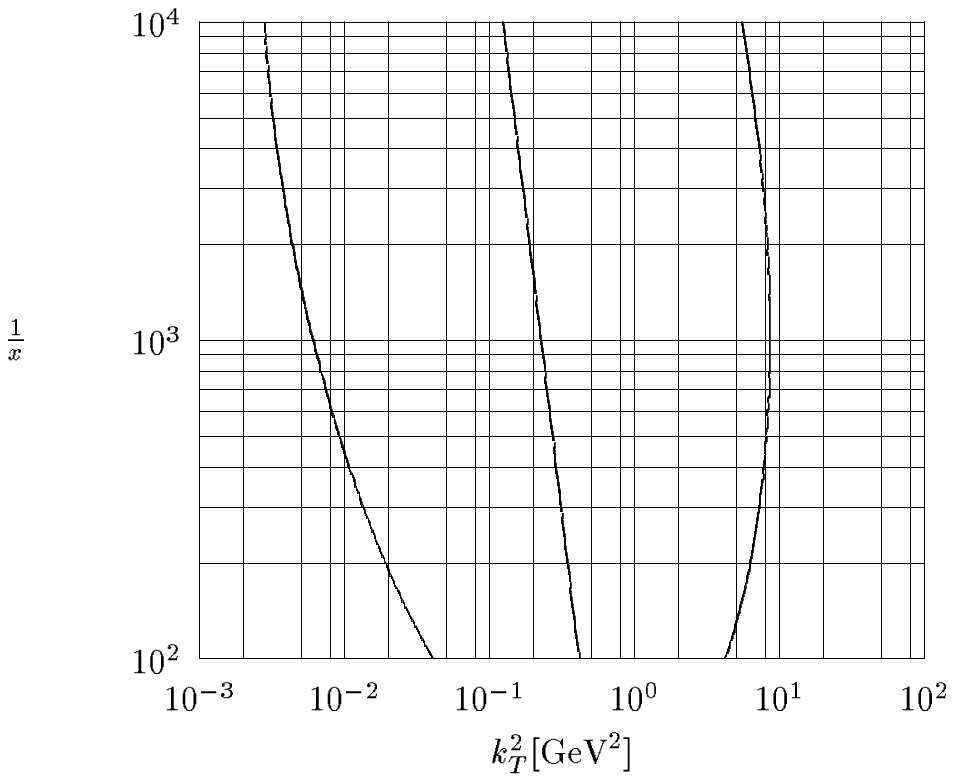,height=31.5cm,width=20.6cm}}
\end{picture}
\end{center}
\caption{}
\label{f4a}
\end{figure}
\begin{figure}[b]
\begin{center}
\begin{picture}(8,8)(0,0)
\put(-6.55,-18.2){\epsfig{file=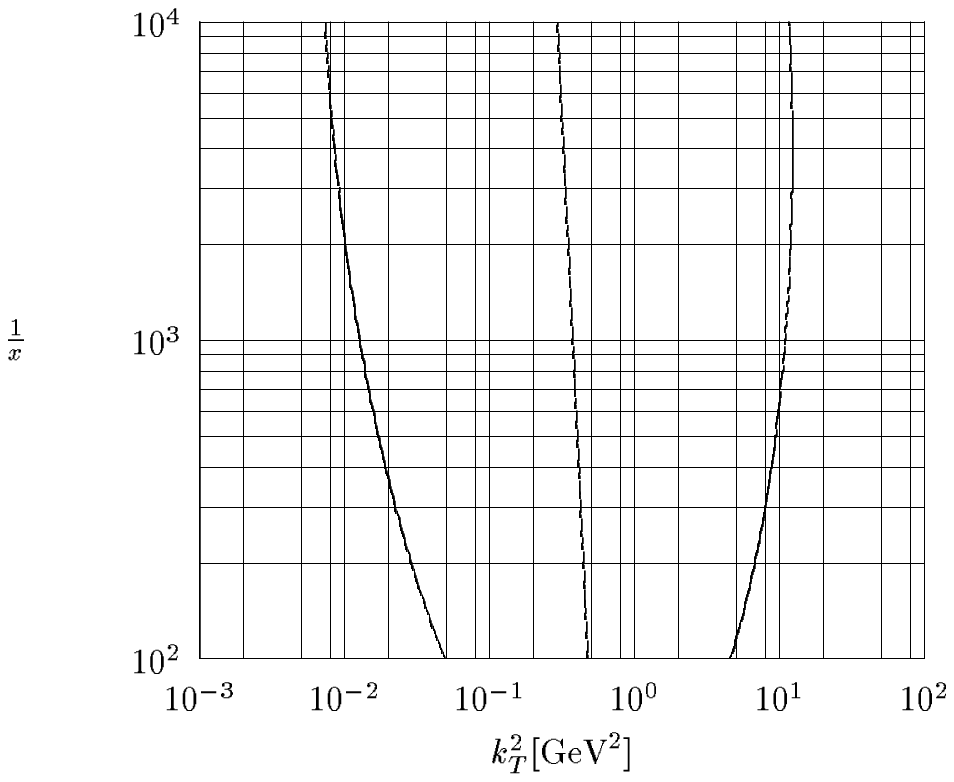,height=31.5cm,width=20.6cm}}
\end{picture}
\end{center}
\caption{}
\label{f4b}
\end{figure}
\clearpage
\newpage
\begin{figure}[t]
\begin{center}
\begin{picture}(8,8)(0,0)
\put(-6.55,-18.2){\epsfig{file=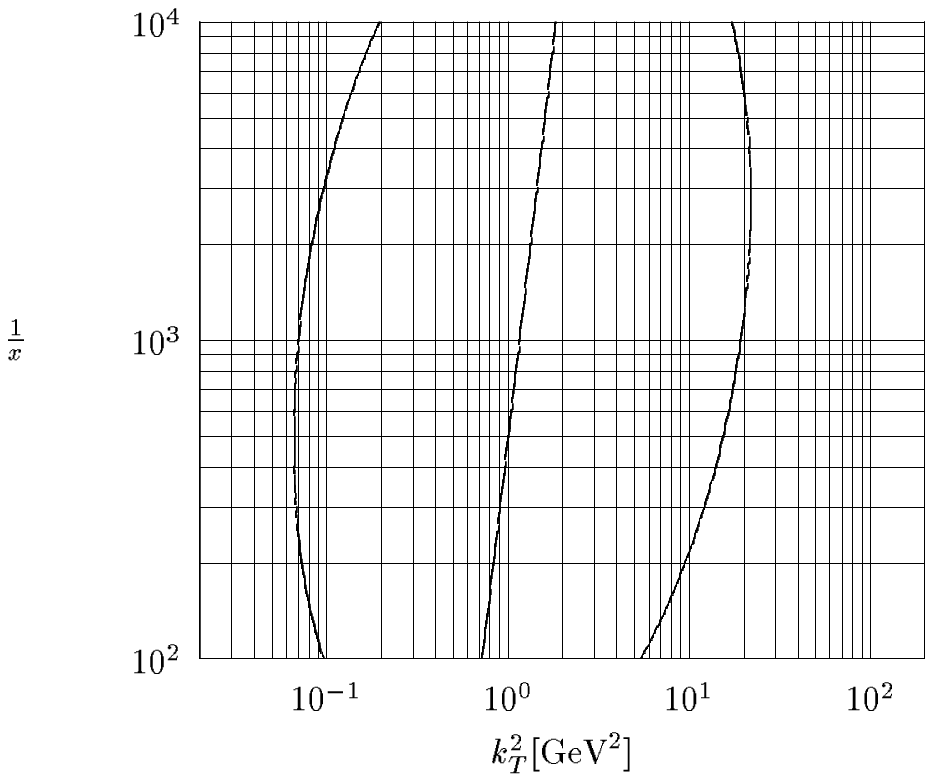,height=31.5cm,width=20.6cm}}
\end{picture}
\end{center}
\caption{}
\label{f4c}
\end{figure}
\begin{figure}[b]
\begin{center}
\begin{picture}(8,8)(0,0)
\put(-6.55,-18.2){\epsfig{file=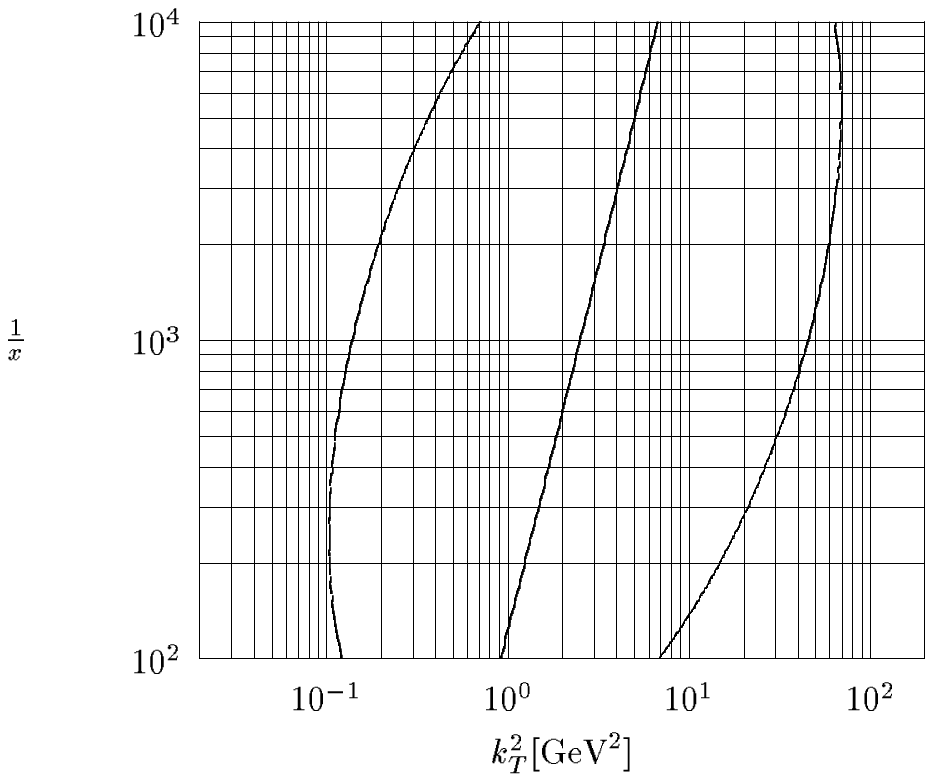,height=31.5cm,width=20.6cm}}
\end{picture}
\end{center}
\caption{}
\label{f4d}
\end{figure}
\clearpage
%
%
\end {document}